\documentclass[conference]{IEEEtran}
\usepackage{graphicx}

\usepackage{amsfonts}
\usepackage{amsthm}

\usepackage[normalem]{ulem}
\usepackage{color}
\usepackage{cite}
\usepackage{amsmath}
\usepackage{subfigure}
\usepackage{footnote}
\usepackage{algorithm}
\usepackage{epstopdf}
\usepackage{array}
\usepackage{amssymb}
\usepackage{enumitem}
\usepackage{epstopdf}
\usepackage{stfloats}
\usepackage{dsfont}  
\usepackage{mdwmath}
\usepackage{mdwtab}
\usepackage{mathtools}
\usepackage{caption}
\usepackage{rotating}

\ifCLASSINFOpdf
\else
\fi
\usepackage{cite}

\begin{document}
\hyphenation{op-tical net-works semi-conduc-tor}
\renewcommand{\thefootnote}{}
\title{Pioneering Studies on LTE eMBMS: Towards 5G Point-to-Multipoint Transmissions}
\author{\IEEEauthorblockN{Hongzhi~Chen\IEEEauthorrefmark{1}, De~Mi\IEEEauthorrefmark{1}, Manuel~Fuentes\IEEEauthorrefmark{2}, David~Vargas \IEEEauthorrefmark{3}, Eduardo~Garro\IEEEauthorrefmark{4}, Jose~Luis Carcel\IEEEauthorrefmark{4}, \\ Belkacem~Mouhouche\IEEEauthorrefmark{2}, Pei~Xiao\IEEEauthorrefmark{1} and Rahim~Tafazolli\IEEEauthorrefmark{1}}
	\IEEEauthorblockA{\IEEEauthorrefmark{1}Institute for Communication Systems, University of Surrey, United Kingdom}
	\IEEEauthorblockA{\IEEEauthorrefmark{2}Samsung Electronics R{\&D} UK, United Kingdom}
	\IEEEauthorblockA{\IEEEauthorrefmark{3}BBC R{\&}D, United Kingdom}
	\IEEEauthorblockA{\IEEEauthorrefmark{4}Institute of Telecommunications and Multimedia Applications, Universitat Politecnica de Valencia, Spain}
	Email:\{hongzhi.chen, d.mi, p.xiao, r.tafazolli\}@surrey.ac.uk, \{m.fuentes, b.mouhouche\}@samsung.com, \\ david.vargas@bbc.co.uk, \{edgarcre, jocarcer\}@iteam.upv.es}


%


\maketitle
\begin{abstract}
The first 5G (5th generation wireless systems) New Radio Release-15 was recently completed. However, the specification only considers the use of unicast technologies and the extension to point-to-multipoint (PTM) scenarios is not yet considered. To this end, we first present in this work a technical overview of the state-of-the-art LTE (Long Term Evolution) PTM technology, i.e., eMBMS (evolved Multimedia Broadcast Multicast Services), and investigate the physical layer performance via link-level simulations. Then based on the simulation analysis, we discuss potential improvements for the two current eMBMS solutions, i.e., MBSFN (MBMS over Single Frequency Networks) and SC-PTM (Single-Cell PTM). This work explicitly focus on equipping the current eMBMS solutions with 5G candidate techniques, e.g., multiple antennas and millimeter wave, and its potentials to meet the requirements of next generation PTM transmissions. \\
\end{abstract}


%
\IEEEpeerreviewmaketitle

{\scriptsize \footnote{We would like to acknowledge the support of the University of Surrey 5GIC (www.surrey.ac.uk/5gic) members for this work. This work was also supported in part by the European Commission under the 5GPPP project 5G-Xcast (H2020-ICT-2016-2 call, grant number 761498). The views expressed in this contribution are those of the authors and do not necessarily represent the project.}}


\section{Introduction}
In the past few years, point-to-multipoint (PTM) technologies have been specified by the 3rd Generation Partnership Project (3GPP) to simultaneously deliver common content to a large amount of users, with a fixed amount of radio resources. 3GPP has enhanced the PTM technologies based on Long Term Evolution (LTE) networks, ever since the adoption of evolved Multimedia Broadcast Multicast Services (eMBMS) in Release (Rel-) 9 \cite{eMBMS}. eMBMS introduces PTM support with small changes on the existing radio and core network infrastructures and protocols of LTE. For instance, new physical, transport and logic channels to enable MBSFN (MBMS over Single Frequency Networks) were added and now coexist with other standardized channels in the specification. Since its introduction, the current eMBMS Rel-14 system has been significantly enhanced and therefore is considerably different from the original version in Rel-9 \cite{D31}. The main novelties introduced up to now include the use of unused MBSFN subframes for unicast reception; MooD (MBMS operation on Demand) which permits to active and deactivate the broadcast service seamlessly and automatically based on the users service consumption reporting; and the introduction of SC-PTM (Single-Cell PTM) to increase the resource allocation flexibility by multiplexing broadcast and unicast data on the same physical channel, instead of using a dedicated multicast channel as MBSFN does. Very recently, a first version of the 5G New Radio Rel-15 specification was approved by 3GPP, where the use cases, the corresponding requirements and technological paradigms are unicast centric or point-to-point (PTP). It is therefore of great interest to extend them to broadcast/multicast-enabling PTM scenarios. Derived from our prior work in \cite{D31}, this work aims at revealing the potential of state-of-the-art PTM technologies previously defined in LTE Rel-14 including the latest fully standardized eMBMS, to meet the requirements of the 5G use cases and scenarios.\\
To this end, we present a comprehensive technical overview of the LTE eMBMS system, and describe its evaluation methodology in detail. A performance analysis and link-level simulations are provided, based on the reference Key Performance Indicators (KPI) and evaluation methodology defined by the ITU-R (International Communications Union - Recommendation) for the IMT-2020 (International Mobile Telecommunication) evaluation process \cite{ITU_R_Guidelines}. Therefore, the results in this work can be served as a benchmark to compare the performance of a potential 5G broadcast solution. The gap analysis of the current eMBMS solutions against the potential requirement of next generation PTM transmissions provides valuable insights into the practical system design (note that the IMT-2020 recommendations are assumed to be a point-to-point/unicast only solution, but we used them as the basis and potential target for evaluating PTM solutions). We also discuss the potential of combining the PTM solutions with 5G candidate techniques, e.g., Massive Multiple-Input Multiple-Output (MIMO) and millimeter wave (mmWave), towards the vertical use cases in next-generation PTM systems.  


\section{Technology Overview}\label{sec:tech_overview}
\begin{figure*}[t]
	\centering
	\includegraphics[width=0.9\textwidth]{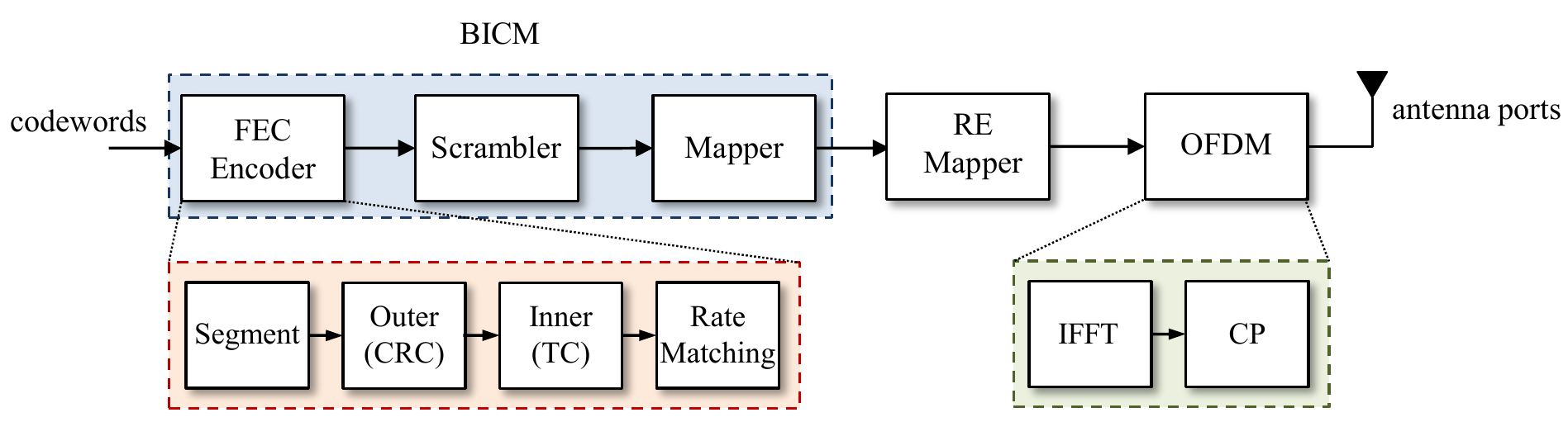}
	\caption{LTE eMBMS physical layer transmitter block diagram}
	\label{blockdiagramembms}
\end{figure*}
The LTE eMBMS Rel-14 physical layer transmitter block diagram is displayed in Fig. \ref{blockdiagramembms}. Input transport block(s) (TB) containing data bits is(are) sent to the Bit-Interleaved Coding and Modulation (BICM) processing chain. Each TB is encoded with a combination of forward error correction (FEC), scrambler and mapper (modulator). More specifically, a Cyclic Redundancy Check (CRC) bit sequence is first attached to each TB. If the TB size is larger than the maximum code block (CB) size, the input data bit sequence is then segmented with an additional CRC sequence attached to each segmented CB. Each code block is then coded using turbo codes (TC) with mother code rate equals to $1/3$. Next, rate matching is performed such that the bits inside each CB are interleaved, circular buffered and punctured/repeated to provide the specific code rate (CR) related to the selected Modulation and Coding Scheme (MCS). Bits after rate matching are then concatenated, scrambled and mapped to constellation symbols. The available modulation schemes for LTE eMBMS are QPSK, 16QAM, 64QAM, and 256QAM. 
After that, symbols are then allocated to all the available resource elements (RE) in the corresponding subframe, and finally, before transmission, CP-OFDM (Cyclic Prefix-Orthogonal Frequency Division Multiplexing) is performed. \\
The eMBMS specification allows for two modes of operation at the physical layer, i.e., MBSFN and SC-PTM, which are introduced in the following subsections
\subsection{MBMS over Single Frequency Network}
The multi-cell solution for LTE broadcasting is called MBSFN, which performs a synchronized transmission of the same content to a group of cells over the same frequency. This group of cells forms an area called Single Frequency Network (SFN) area. Compared to SC-PTM, MBSFN eliminates inter-cell interference within the whole service area. 
MBSFN provides PTM services using a specific Physical Multicast Channel (PMCH) and occupying the entire bandwidth. Regarding the subcarrier spacing, MBSFN can be configured with values of 15, 7.5 and 1.25 kHz, corresponding to extended CP durations of 16.7, 33.3 and 200 {$\mu$}s respectively. Since a separate physical channel is used, MBSFN employs different reference signals compared to LTE unicast and SC-PTM. eMBMS is standardized to
multicast data over a single antenna port, and therefore it does not take advantage of the potential MIMO capability, thus only a single codeword can be sent.
\subsection{Single Cell Point to Multipoint}
SC-PTM is another LTE PTM solution which aims at providing broadcasting service to groups of users in one single cell. To that end, the resource allocation flexibility is increased by employing the Physical Downlink Shared Channel (PDSCH), which is also used for unicast transmission. It should be noted that sharing a physical channel implies to use the same carrier spacing for both technologies. As a consequence, only 15 kHz subcarrier spacing with normal CP is available for SC-PTM. Since different contents are transmitted in neighbour cells, transmission power management is required in order to control inter-cell interference, 
\section{Evaluation Methodology and Analysis}\label{sec:methodology}
The technical performance requirements about the primary usage scenarios and their corresponding evaluation methodology for PTP transmissions are defined in the IMT-2020 evaluation process \cite{ITU_R_Requirements}. In addition, the methodology for the requirement we consider in this paper also follows these guidelines \cite{ITU_R_Guidelines}. 
{Following the same methodology, this study is focused on BICM spectral efficiency (SE) that is considered in the IMT-2020 KPI set. This is because the BICM SE provides the capacity supported by PTM solutions for different modulation and coding combinations.}
\subsection{BICM Spectral Efficiency Evaluation}
BICM spectral efficiency denotes the spectral efficiency that can be achieved by the PTM system while using different modulation schemes, excluding the resource that are used for control information and synchronisation, i.e., overheads are not taken into account. The BICM spectral efficiency is expressed in bits per RE (bits/RE) and can be calculated by:

\begin{equation}
\text{SE}_{\text{\emph{BICM}}} = \log_2(M) \cdot \text{CR} \cdot N_{\emph{Tx/Rx}}
\label{Eq5}
\end{equation}

\noindent where $M$ denotes the modulation order, i.e., for QPSK, $M=4$, $\text{CR}$ represents the effective code rate corresponding to the MCS index and $N_{\emph{Tx/Rx}}$ is the number of independent information spatial streams with multiple transmitter and receiver antennas. Note that the effective CR in LTE is calculated as:

\begin{equation}
\text{CR} = \frac{\text{TBS}}{N_{\emph{avail}}}
\label{Eq6}
\end{equation}

\noindent where $\text{TBS}$ represents the transport block size, which is fixed for each modulation and coding index, can be found in \cite{TR36_213}, and $N_{\emph{avail}}$ is the number of total available bits in one subframe, given by:

\begin{equation} 
N_{\emph{avail}} = m \cdot N_{\text{\emph{RB}}} (N_{\emph{sym}}N^{\text{\emph{RB}}}_{sc} - N_{\emph{RS}})
\label{Eq7}
\end{equation}

\noindent For \eqref{Eq7}, $m=\log_2(M)$, denotes the number of bits per constellation symbol, $N_{\text{\emph{RB}}}$ is the number of resource blocks (RB) utilized within a subframe, $N_{\emph{sym}}$ is the number of OFDM symbols per subcarrier excluding those used for control channel, $N^{\text{\emph{RB}}}_{sc}$ is the number of subcarriers per RB, and $N_{\emph{RS}}$ is the number of reference signals per RB.
Note that the maximum CR cannot exceed 0.925, which is the CR associated to the maximum Channel State Indicator (CQI) 15. The maximum TB size with SC-PTM and MBSFN is given for indexes $I_{\text{\emph{TBS}}} = 33$ and $32$ \cite{TR36_213} and the associated CR is then 0.887 and 0.882 respectively.
Also, the required Carrier to Noise Ratio (CNR) to achieve a given BICM SE value depends on the channel environment. 

\subsection{BICM Spectral Efficiency Calculation}
\begin{figure}[t]
	\centering
	\includegraphics[width=0.49\textwidth]{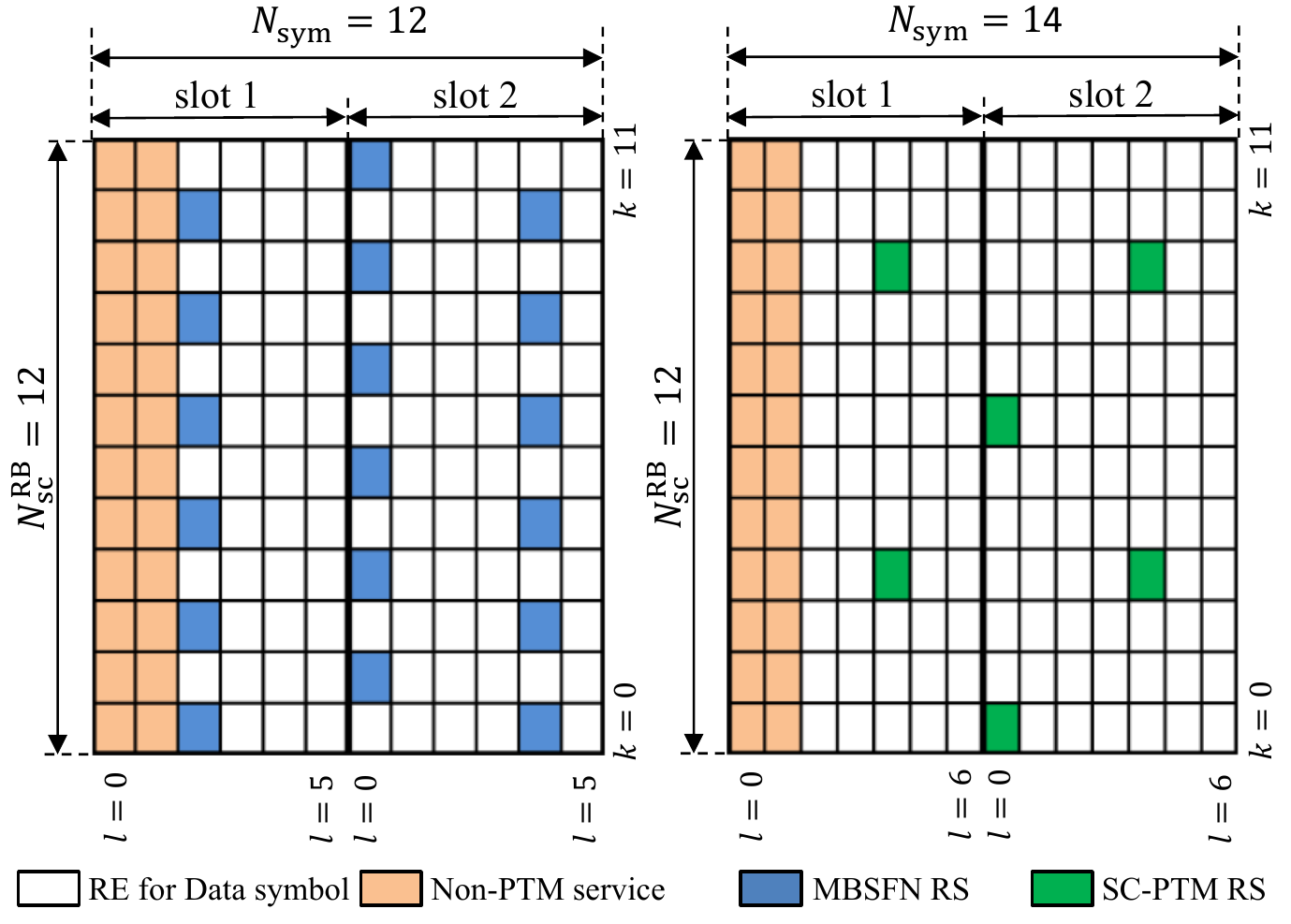}
	\caption{Resource Block frame structure for MBSFN (left) and SC-PTM (right), with 15 kHz carrier spacing.}
	\label{Fig_Framestructure}
\end{figure} 
As previously mentioned, the BICM spectral efficiency depends on the modulation order and effective code rate, each Modulation and Coding Scheme (MCS) provides a different CR directly related to a particular TB size.
Fig. \ref{Fig_Framestructure} shows the different frame structure for a single RB for both configurations, i.e. MBSFN and SC-PTM \cite{TR36_213}. Note that the carrier spacing selected is 15 kHz for illustration purposes, and the number of OFDM symbols per subcarrier with MBSFN and SC-PTM is 12 and 14, due to the use of extended and normal CP respectively. As Fig. \ref{Fig_Framestructure} depicts, the number of subcarriers per RB with MBSFN and SC-PTM are both 12.
In SC-PTM, the number of OFDM symbols per subcarrier used for the control channel is 3 for 5 MHz bandwidths and below and 2 otherwise. The available bits with SC-PTM when considering 50 RBs, can be calculated using (\ref{Eq7}), obtaining $m \cdot 50(12\cdot12 - 6) = 6900m$ bits,  considering a channel bandwidth of 10 MHz. In MBSFN, 3 different configurations can be used, with different carrier spacing. In this paper, studies for MBSFN are focused on the standalone mode with carrier spacing 1.25 kHz. In this case, the available bits $N_{\emph{avail}} = m\cdot50(1\cdot144 - 24) = 6000m$ bits.
Since the maximum constellation size availbale is 256QAM, the peak value of BICM spectral efficiency is 7.09 and 7.06 bits/RE with SC-PTM and MBSFN respectively. Note that the same calculation can be easily extended to MIMO. SC-PTM with 4 spatial streams (MIMO $4\times4$) can reach up to 28.36 bits/RE, whereas MBSFN is limited to 7.06 bits/RE since the use of MIMO is not specified.

\section{Link-Level Simulation Evaluation}\label{sec:performance}
Link-level results for the BICM spectral efficiency as a function of the required CNR for SC-PTM and MBSFN are presented in this section. Different scenarios i.e., channel types have been evaluated in order to assess the impact of the configurations adopted. The selected Quality of Service (QoS) metric is block error rate (BLER) lower than 0.1\%.  A subcarrier spacing of $\Delta_f = 1.25$ kHz is always used with MBSFN, in order to study the potential advantages of this mode compared to SC-PTM ($\Delta_f = 15$ kHz). The simulation parameters we considered in this paper are aligned with the IMT-2020 evaluation guidelines \cite{ITU_R_Guidelines}.
\subsection{Additive White Gaussian Noise Channel}
\begin{figure}[t]
	\centering
	\includegraphics[width=0.49\textwidth]{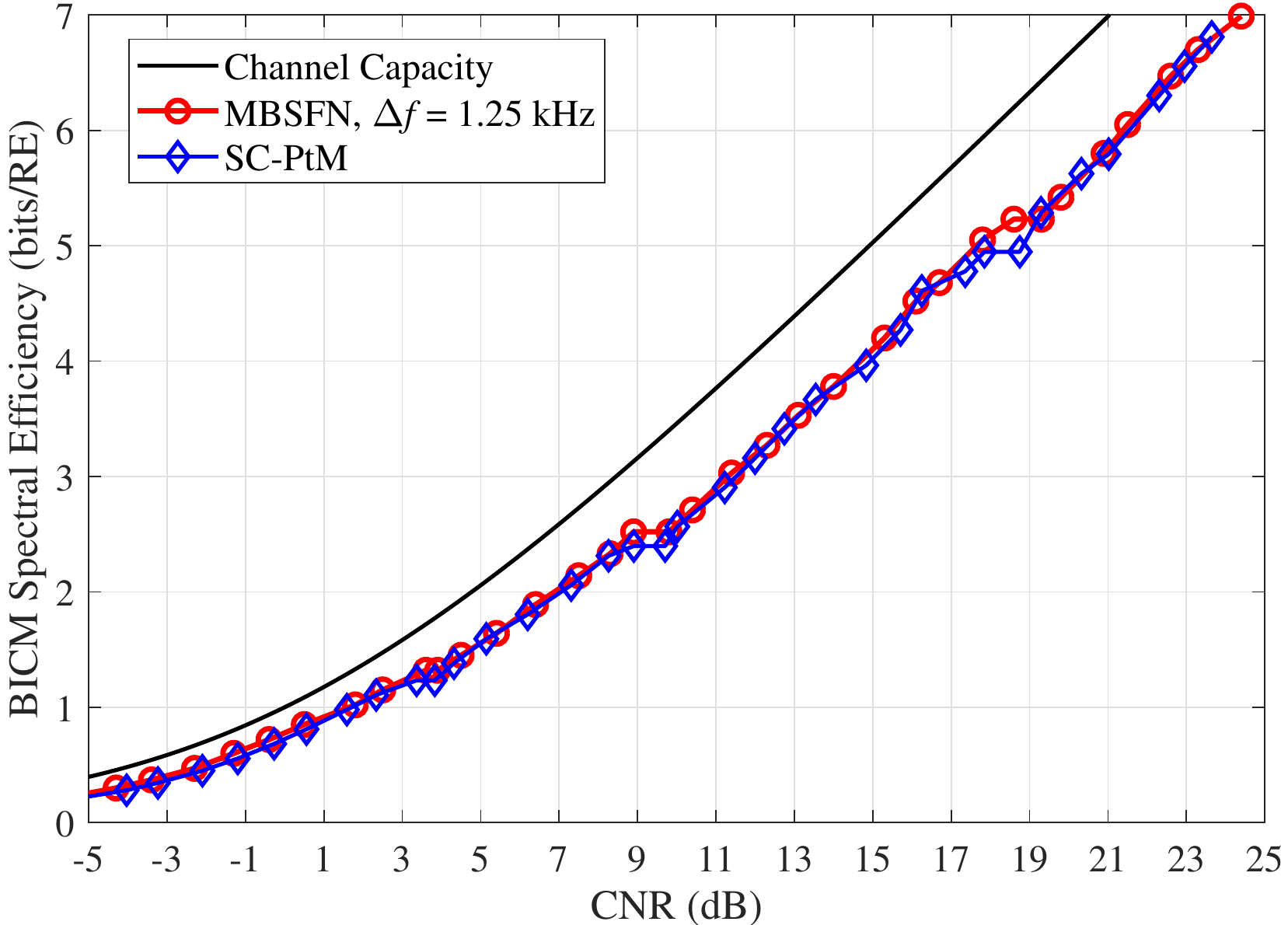}
	\caption{BICM SE vs. CNR (dB) for AWGN channel.}
	\label{Fig_Performance_AWGN}
\end{figure} 
Fig. \ref{Fig_Performance_AWGN} depicts the comparison of the two PTM schemes in Additive White Gaussian Noise (AWGN) channel as well as the channel capacity. Both transmitter and receiver are considered equipping single antenna. Although SC-PTM and MBSFN are both PTM solutions, we consider a simplified model where one user inside one single cell is simulated.
Note that this simplified scenario may present the considered PTM solution in its best light. This can be expressed from two aspects: 1) The channel been simulated is an AWGN channel; 2) The data rate of a PTM system depends on the weakest user's data rate, which means that a PTP case represents the best scenario in terms of the achievable data rate of the system. From the results, we can see that the overall BICM SE follows the same trend regardless of the PTM technology used, and the clearly visible gap between the achievable BICM SE and channel capacity can be observed.
\begin{figure}[t]
	\centering
	\includegraphics[width=0.45\textwidth]{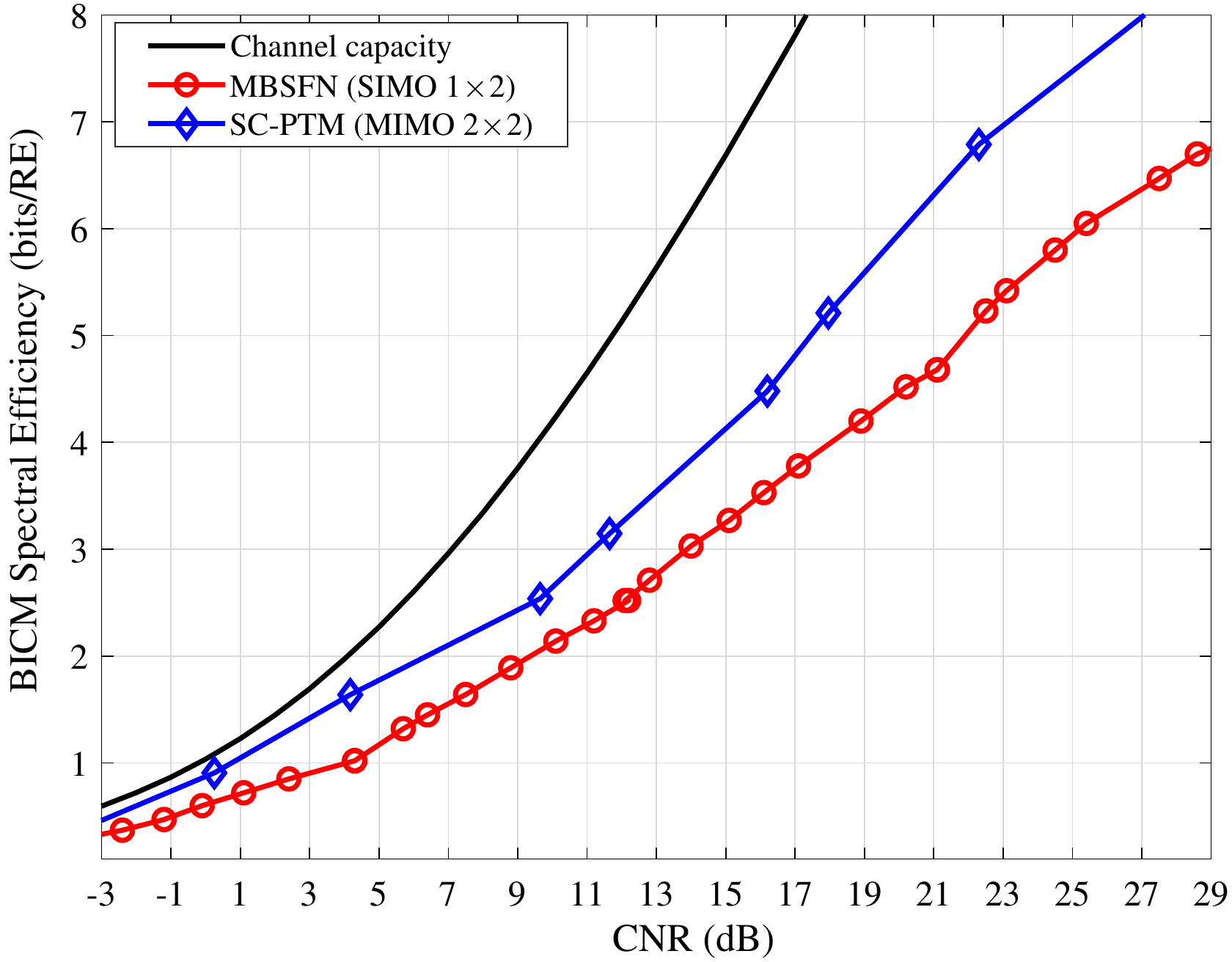}
	\caption{BICM SE vs. CNR (dB) for MIMO/SIMO i.i.d Rayleigh channel.}
	\label{Fig_MIMO}
\end{figure} 
\subsection{Independent Identical Distributed Rayleigh Channel}
Fig.~\ref{Fig_MIMO} shows the BICM spectral efficiency for the independent and identically distributed (i.i.d.) Rayleigh MIMO channel.  In this simulation, SC-PTM utilizes 2-by-2 MIMO, while MBSFN employs a $1\times2$ Single-Input Multiple-Output (SIMO) scheme. Minimum Mean Square Error (MMSE) demapper has been used with SC-PTM in order to cope with computation complexity limitations with higher modulation orders. For SC-PTM, representative MCS indexes from 0 to 27 with step 3 are used.
{From Fig.~\ref{Fig_MIMO}, spatial multiplexing of multiple codewords, i.e., SC-PTM with MIMO, provides a higher BICM SE compare to the diversity gain offered by multiple receive antennas with only one transmitted codeword, e.g., MBSFN with SIMO, and the gap increases at the high CNR region.} 
As calculated in section.~\ref{sec:methodology} the BICM SE for MBSFN is limited to 7 bits/RE, SC-PTM can increase its limit to 9 bits/RE, respectively.


\section{Discussion: PTM Potential Enhancements }\label{sec:comments}
\subsection{Combining MBSFN with Smart Antenna Techniques}
In Fig.~\ref{Fig_MIMO}, it can be seen that the performance gap between multiplexing multiple codewords and diversity transmission/receiving of a single codeword is significant, especially in the high CNR regime. This indicates that enabling MIMO techniques in MBSFN scenarios have potential to increase the system throughput/SE dramatically. Combining MBSFN with MIMO provides diversity transmission and multiplexing. It is then possible to assume that we have an SFN area comprised of $M$ cells, where each cell has one base station (BS) equipped with two transmit antennas. Thus the same codewords are multiplexed onto the two transmit antennas in each cell, and therefore each of the $M$ BSs can broadcast the multiplexed signal to all users inside the SFN area. Assuming there are $N$ single-antenna users in the SFN area, we can form a 2M-by-N distributed MIMO scheme. When $M$ goes larger. i.e., in the case of a increased SFN area, the considered system can be categorized as massive MIMO \cite{MMNG}. It is worth mentioning that, the bigger the SFN area is, the harder to perform synchronized transmission. Also the frame structure need to be adopted, e.g., CP length, subcarrier spacing. 
\subsection{Operating PTM solution in Millimetre Wave band}
To meet the requirement of new generation PTM transmissions with ultra high definition video streaming, large bandwidth is preferred to provide over Gbps bit rates and overcome the spectrum scarcity problem \cite{ITU_R_Requirements}. In this context, the application of mmWave communications has been shown as a potential promising solution, where a large amount of underutilized band can be leveraged to provide the potential gigahertz transmission bandwidth \cite{mmWaveItWillWork}. However, the extreme narrow beam formed in the traditional mmWave antenna array design is in general not suitable for broadcasting or multicasting. Novel designs on high-gain omnidirectional slotted-waveguide antenna arrays have been recently proposed \cite{SWAAC1NP}, whose radiation pattern shows the promising coverage in an Olympic-standard arena by using only one high-gain radiating element. This can fundamentally change the design principle of the next generation PTM transmissions and bring new solutions, e.g., SC-PTM in small cells with 100-fold boosted capacity based on the existing standardized PTM technologies. 

\section{Conclusion}\label{sec:conclusion}
In this paper, the state-of-the-art LTE eMBMS PTM technologies, i.e., MBSFN and SC-PTM, have been analysed. The detailed technical overview of both MBSFN and SC-PTM and their designs have been covered. The analysis presented, based upon the link-level simulations has revealed that regarding BICM spectral efficiency, 
SC-PTM equipped with MIMO can provide up to four times more capacity (bits/RE) than MBSFN, since the last one does not support the use of multiple antennas with multiple codewords. This work also discussed the potential of combining MBSFN and SC-PTM with 5G candidate techniques such as massive MIMO and mmWave, serving as a useful guidance to improve the current LTE eMBMS PTM technologies.

Further investigations on comparing the LTE eMBMS with other terrestrial broadcasting standards are also needed, in order to provide a comprehensive gap analysis on current PTM technologies. 






%
\bibliographystyle{IEEEtran}
\bibliography{reference}

\end{document}